\documentclass[]{aa}
\usepackage[varg]{txfonts}
\usepackage{lineno}
\usepackage{natbib}
\usepackage{amsmath}
\usepackage{graphicx}
\usepackage{wasysym}
\usepackage{txfonts}
\usepackage{xspace}
\usepackage{url}
\usepackage{textcomp}
\usepackage{multirow}
\usepackage{lipsum}

\newcommand{\exo}{EXO\,2030+375\xspace}

\newcommand{\swift}{\textsl{Swift}\xspace}
\newcommand{\fermi}{\textsl{Fermi}\xspace}

\newcommand{\inte}{\textsl{INTEGRAL}\xspace}

\newcommand{\xmm}{\textsl{XMM-Newton}\xspace}

\newcommand{\suz}{\textsl{Suzaku}\xspace}

\newcommand{\nustar}{\textsl{NuSTAR}\xspace}

\newcommand{\asec}{\ensuremath{''}\xspace}

\newcommand{\snr}{S/N\xspace}

\newcommand{\msun}{\ensuremath{\text{M}_{\odot}}\xspace}

\newcommand{\redchi}{\ensuremath{\chi^{2}_\text{red}}\xspace}

\newcommand{\feka}{\ensuremath{\mathrm{Fe}~\mathrm{K}\alpha}\xspace}

\newcommand{\ha}{\ensuremath{\mathrm{H}\alpha}\xspace}

\newcommand{\nh}{\ensuremath{{N}_\mathrm{H}}\xspace}

\newcommand{\nhone}{\ensuremath{{N}_{\mathrm{H},1}}\xspace}
\newcommand{\nhtwo}{\ensuremath{{N}_{\mathrm{H},2}}\xspace}

\newcommand{\ergcms}{\ensuremath{\text{erg\,cm}^{-2}\text{s}^{-1}}}

\newcommand{\tone}{Type~I\xspace}
\newcommand{\ttwo}{Type~II\xspace}

\begin{document}

\title{Studying the accretion geometry of \exo at luminosities close to the propeller regime}

\author{F.~F\"urst\inst{1}\and P.~Kretschmar\inst{1}\and J.~J.~E.~Kajava\inst{2,3,1}\and J.~Alfonso-Garz\'on\inst{4}\and M. K\"uhnel\inst{5}\and C. Sanchez-Fernandez\inst{1}\and P. Blay\inst{6,7}\and C.~A.~Wilson-Hodge\inst{8}\and P.~Jenke\inst{8}\and   I.~Kreykenbohm\inst{5}\and K.~Pottschmidt\inst{9,10}\and  J.~Wilms\inst{5}\and R.~E.~Rothschild\inst{11}}

\offprints{F.~F\"urst, \email{felix.fuerst@sciops.esa.int}}

\institute{European Space Astronomy Centre (ESAC), Science Operations Departement, 28692 Villanueva de la Ca\~nada, Madrid, Spain
  \and Finnish Centre for Astronomy with ESO (FINCA), University of Turku, V\"{a}is\"{a}l\"{a}ntie 20, 21500 Piikki\"{o}, Finland
  \and Tuorla Observatory, University of Turku, V\"{a}is\"{a}l\"{a}ntie 20, 21500 Piikki\"{o}, Finland
  \and Centro de Astrobiolog\'ia - Departamento de Astrof\'isica (CSIC-INTA), Camino Bajo del Castillo s/n, Urb. Villafranca del Castillo, 28691 Villanueva de la Ca\~nada, Madrid, Spain
  \and Dr. Karl-Remeis-Sternwarte and ECAP, Sternwartstr. 7, 96049 Bamberg, Germany
  \and Instituto de Astrof\'isica de Canarias, Tenerife, Spain
  \and Nordic Optical Telescope, La Palma, Spain
  \and NASA Marshall Space Flight Center, Huntsville, AL 35812, USA
  \and CRESST, Department of Physics, and Center for Space Science and Technology, UMBC, Baltimore, MD 21250, USA
  \and NASA Goddard Space Flight Center, Greenbelt, MD 20771, USA
  \and Center for Astrophysics and Space Sciences, University of California, San Diego, 9500 Gilman Dr, La Jolla, CA 92093-0424, USA
  } 

\date{Received XX.XX.XX / Accepted XX.XX.XX}

\abstract{
The Be X-ray binary \exo was in an extended low-luminosity state during most of 2016. We observed this state with \nustar and \swift, supported by \inte observations and  optical spectroscopy with the Nordic Optical Telescope (NOT). We present a comprehensive spectral and timing analysis of these data here to study the accretion geometry and investigate a possible onset of the propeller effect. 
The H$\alpha$ data show that the circumstellar disk of the Be-star is still present. We measure equivalent widths similar to values found 
during more active phases in the past, indicating that the low-luminosity state is not simply triggered by a smaller Be disk.
The \nustar data, taken at a 3--78\,keV luminosity of  $\sim$$6.8\times10^{35}$\,erg\,s$^{-1}$ (for a distance of 7.1\,kpc), are nicely described by standard accreting pulsar models such as an absorbed power law with a high-energy cutoff. We find that pulsations are still clearly visible at these luminosities, indicating that accretion is continuing despite the very low mass transfer rate.
In phase-resolved spectroscopy we find a peculiar variation of the photon index from $\sim$1.5 to $\sim$2.5 over only about 3\% of the rotational period.  This variation is similar to that observed with \xmm at much higher luminosities. It may be connected to the accretion column passing through our line of sight.
With \swift/XRT we observe luminosities as low as $10^{34}$\,erg\,s$^{-1}$ {where the data quality did not allow us to search for pulsations, but the spectrum is much softer and well described by either a blackbody or  soft power-law continuum. This softer spectrum might} be due to the accretion being stopped by the propeller effect and we only observe the neutron star surface cooling. }


\maketitle

\section{Introduction}
\label{sec:intro}
Transient, accreting neutron stars in Be X-ray binaries can show luminosity variations over two to three orders of magnitude or more \citep[e.g.,][and references therein]{negueruela98a}. This variability allows us to investigate the X-ray producing region in the accretion column over a wide variety of physical conditions, i.e., over regimes in which different interaction processes between the infalling material and the X-ray radiation are dominant.

An open question is if and how accretion takes place at the lowest luminosities. Here, the strong magnetic field of the neutron star might inhibit further accretion, in what is known as the  ``propeller regime'' \citep{illarionov75a}.  In this regime, the magnetospheric radius becomes larger than the  co-rotation radius, due to the reduced ram pressure of lower density material surrounding the neutron star. This effect inhibits accretion, leading to a drastically reduced accretion luminosity. Any observed residual flux might then originate from the hot neutron star surface or weak residual accretion at the poles or might be due to instabilities at the magneotsphere. 

On the other hand, \citet{tsygankov17a} have shown that for slow rotators a cold accretion disk could form, preventing the source from entering the propeller regime. Obtaining a better understanding of if and  how accretion takes place at very low luminosities will help  inform accretion modeling and allows us to study  the magnetospheric interactions in greater detail. However,  the intrinsic low fluxes make this regime  difficult to observe.

One of the best studied Be X-ray binaries is \exo, a neutron star in orbit with a B0\,Ve star \citep{reig98a}  at a distance of about 7.1\,kpc \citep{wilson02a}. Since its discovery in 1985 \citep{coe88a,parmar89a}  it has shown outbursts very regularly close to each periastron passage of its 45\,d orbit, which are called \tone outbursts. They occur when the eccentric orbit of the neutron star brings it close enough to the circumstellar disk of its companion to trigger enhanced mass transfer. \tone outbursts are therefore significantly shorter than the orbital period,  typically with durations between 10 and 20\,days. In \exo over 150 \tone outbursts have been observed \citep{laplace17a}.

In addition to the many \tone outbursts, \exo showed two much larger  \ttwo outbursts in 1985 and 2006, that lasted for multiple orbital cycles. These events reach much higher luminosities and are likely due to increased activity of the Be-star and complex interaction between the neutron star and the circumstellar disk. These interactions lead to instabilities in which the mass transfer onto the neutron star is further increased \citep[e.g.,][]{okazaki01a}. These spectacular outbursts were intensely monitored and led to insights about the high-luminosity accretion regime   \citep{parmar89b, parmar89a, reynolds93a, klochkov07a, wilson08a}. For example, \exo was the first pulsar in which a dramatic change in the pulse profile was observed as a function of luminosity, indicating a change in the emission pattern of the accretion column \citep{parmar89a}.

\citet{ferrigno16a} studied phase-resolved \xmm and \suz data of \exo during \tone outbursts in detail. They found a unique feature in the pulse profile, where the spectral hardness varies drastically over only about 2--3\% of the rotational period. They interpret this feature as part of the accretion column passing through our line of sight towards the main X-ray producing region, resulting in a higher absorption column and in a more reprocessed spectrum. \citet{ferrigno16a} argue that the accretion column  grows in size with increasing X-ray luminosity and therefore the duration of the feature should also increase. However, due to the limited phase-space resolution possible with \suz, they could not confirm this theory using a brighter \suz observation.

While  \tone outbursts always occur close to periastron passage, their exact orbital phase has varied with time in \exo. In particular, \citet{wilson02a} observed that in late 1995 the outbursts occurred a few days before periastron, instead of after, but were overall fainter compared to previous outbursts. Around that time the source also switched to a long-term spin-down of its 42\,s pulse period, the first such trend observed for a source that was still showing regular outbursts \citep{wilson02a}. 
 
 \citet{wilson05a} found that in 2003 \exo switched back to a spin-up trend and that the outbursts became brighter again, which led them to postulate a roughly 11-year cycle of the activity of \exo.  By comparing the complete history of the pulse period and the outburst phase between 1985 and 2016, \citet{laplace17a} also find an 11-year periodicity; however, these authors also suggest that the true period might be about twice as long, i.e., 21\,years. On this period, giant outbursts alternate with a series of faint outbursts, explained by the alternating maxima of eccentricity and inclination of the Be-disk due to the Kozai-Lidov effect \citep{martin14a}. In early 2015 the regularity of the outbursts started to break up again and an orbital phase shift was observed in 2016 July, which  seems to confirm this 21-year period \citep[see also][]{atel9263}

During the start of the faint phase in 2015  an outburst was observed only at every other periastron passage. Soon after, X-ray monitors like \swift/BAT, MAXI, and \fermi/GBM could not detect any activity \citep{atel8835}. 
This period of quiescence provided the ideal opportunity to study \exo at the lowest possible luminosities to search for evidence of changes in the accretion geometry and for the onset of the propeller effect. We triggered a director's discretionary time target-of-opportunity observation with the \textsl{Nuclear Spectroscopic Telescope Array} (\nustar) as well as a longer \swift observation campaign to follow several periastron passages.

As noted by \citet{atel9485} and \citet{laplace17a} the outbursts of \exo picked up again by the end of 2016. 
Since then, the peak flux of the \tone outbursts occurs around orbital phase $\phi=0.0$, based on the ephemeris of Fit~2 by \citet{wilson08a}. This phase of maximum flux is significantly earlier in phase than the regular outbursts before 2015, which occurred closer to $\phi=0.1$, i.e., about 4--5 days later \citep{kretschmar17a}. At the time of writing, \exo appears to have entered a more regular outburst regime, with sequential outbursts from  December 2016 to March 2017. All these outbursts also peak at the earlier phase, around $\phi=0.0$.

In this paper we present timing and spectral analysis of  \nustar and \swift data, as well as an analysis of serendipitous observations with the \textsl{International Gamma-Ray Astrophysics Laboratory} (\inte) and optical spectroscopy with the Nordic Optical Telescope (NOT) taken during the quiescence period of 2016. In Sect.~\ref{sec:data} we describe the data reduction. In Sect.~\ref{sec:spec} we perform spectral analysis  of the \nustar observation. In Sect.~\ref{sec:longterm} we put these \nustar results into context and describe the long-term flux and timing evolution. We  discuss our results  in Sect.~\ref{sec:disc} and present our conclusions in Sect.~\ref{sec:concl}.

\section{Observations and data reduction}
\label{sec:data}

All X-ray data reduction was performed with HEASOFT v6.19. Data analysis was performed with the Interactive Spectral Interpretation System \citep[ISIS,][]{houck00a} v1.6.2. All uncertainties are reported at the 90\% confidence level for one parameter of interest unless otherwise noted.

\subsection{\nustar}
\label{susec:nustar}

\nustar \citep[][]{harrison13a} observed \exo starting on 2016 July 25, 08:36 UTC (ObsID  90201029002). We extracted the data using the standard software \texttt{nupipeline} v1.6 and CALDB v20160606. To increase the usable exposure time, we also extracted  SCIENCE\_SC data (mode 06), following the procedures as discussed by \citet{walton16a} and \citet{gx339IHS}. This approach resulted in an exposure time of 51\,ks for each module.

The source spectra were extracted separately for each focal plane module and data mode, using a circular region with 40\asec radius centered on the source image in the respective detector. The region size was chosen to optimize the signal-to-noise ratio (S/N) at the highest energies: a larger region adds more background than source photons above 60\,keV. Background spectra were extracted from a circular region with 120\asec radius on the same quadrant of the detector as the source, avoiding all visible stray light.

\subsection{\swift}
\label{susec:swift}

We extracted  data from the X-ray Telescope \citep[XRT,][]{swiftxrtref} from all 31 \swift \citep{swiftref} pointings between 2016 March and 2016 October  using our custom scripts based on the standard XRT extraction method using \texttt{xrtpipeline}, \texttt{xselect}, and \texttt{xrtmkarf}. All observations were performed in photon-counting mode and have a typical exposure time of about 1\,ks. We mostly used a circular source extraction region with a radius of 30$''$; however, for the bright observations we used an annulus region with an inner radius of 15$''$ and outer radius of 45$''$ to avoid pile-up. The background was extracted from a circular region with 200$''$ radius.

\subsection{\inte}
\label{susec:inte}

\exo was serendipitously observed by \inte \citep{integralref} during a
series of public observations of Cyg~X-1 scheduled for 2016 October 27--31
 (Prop.~ID 1220044). The \inte IBIS/ISGRI instrument \citep{isgriref}  provided complementary hard X-ray data to the soft X-ray coverage of \swift.
Given the moderate source brightness over
this period, we filtered the available data set and selected only
those pointings for which the source was detected within the
$8\fdg3\times8\fdg0$ Fully Coded Field of View of IBIS/ISGRI.  This selection resulted in a net exposure time of 194\,ks. The IBIS/ISGRI data reduction was performed using the
Off-line Scientific Analysis software (OSA; \citealt{isdcref}) v10.2,
using the latest calibration files. Following standard reduction
procedures the data were processed from the
COR step to the SPE step .

\subsection{Nordic Optical Telescope}
\label{susec:optical}
The NOT is a 2.5\,m optical telescope located on the Observatorio del Roque de los Muchachos, La Palma, Canarias. We took spectra with its ALFOSC spectrograph during the nights of 2016 June 14, July 31, August 21, and September 21. ALFOSC was equipped with a e2v CCD,  with 2048$\times$2064 pixel size. The instrumental setup included a 1\asec slit and a 485 lines per millimeter volume phase  holographic grism, with a central wavelength of 7850\,\AA, a dispersion of 2.2\,\AA~per pixel, and a resolution of 770, covering the wavelength range 5650--10150\,\AA. The data were reduced with IRAF \citep{irafref}, applying standard procedures for long-slit spectroscopy.

\section{\nustar spectroscopy}
\label{sec:spec}
\subsection{Phase-averaged spectra}

To compare the \exo low-luminosity spectrum to previous observations, we fit the phase-averaged \nustar data together with almost simultaneous \swift/XRT data (ObsID 00030799022) with typical phenomenological models often used to describe the X-ray spectra of these systems. We rebinned the \nustar data within ISIS to a \snr of 6 below 45\,keV and 3 above, while requiring  a rebinning factor of at
least 2. This approach allows us to use the data between 3--70\,keV.  The XRT data were rebinned to a \snr of 3 throughout, which provided useful data between 1--8\,keV. The XRT data were taken a bit later in the outburst where the flux already started declining, hence showing a lower flux than the \nustar data, which we model via a multiplicative constant. However, the changes in spectral shape over this time range are negligible. 

\exo is known to show complex absorption that is often fitted with a partially covering absorber \citep[see, e.g.,][]{naik13a,ferrigno16a}. We modeled the data using a global absorption column, \nhone, and an additional partially covering column, \nhtwo, with a covering fraction $f$. To describe the absorption we used an updated version of the \texttt{tbabs} model\footnote{\url{http://pulsar.sternwarte.uni-erlangen.de/wilms/research/tbabs/}} by \citet{wilms00a}. We used the abundance provided by these authors and the cross sections by \citet{verner96a}. In ISIS this model can  be expressed as 
\begin{equation}
\nhone \times \left[f  \nhtwo + \left(1-f\right)\right] \times \text{continuum}
.\end{equation}

As continua we tried \texttt{cutoffpl}, \texttt{highecut}, \texttt{NPEX} \citep{mihara98a}, and \texttt{FDcut} \citep{tanaka86a}. The results are listed in  Table~\ref{tab:phasavg}. A summary of the definitions of each of these models can be found in \citet{mueller13a}.  We find that neither the \texttt{cutoffpl} nor the \texttt{FDcut} provided a statistically acceptable fit to the data, while both \texttt{highecut} and \texttt{NPEX} resulted in very good and statistically equivalent fits. For the \texttt{highecut} model we also show the results without the partially covering absorber in Table~\ref{tab:phasavg}, which is significantly worse, with an F-test false detection probability of $3.7\times10^{-5}$.

\begin{table*}
\caption{Best-fit model parameters for the phase-averaged fits. PC denotes models with partial covering, while  $CC_\text{B}$ and  $CC_\text{XRT}$ are the cross-calibration constants for \nustar/FPMB and \swift/XRT, respectively, with respect to \nustar/FPMA.}
\label{tab:phasavg}
\centering
\begin{tabular}{rllllll}
\hline\hline
 Parameter & HighE (no PC) & HighE (with PC) & HighE+bbody & NPEX & FDcut & CPL \\\hline
 $ N_\text{H,1}~(10^{22}\,\text{cm}^{-2})$ & $6.1\pm0.7$ & $4.1^{+1.1}_{-2.1}$ & $4.0\pm0.7$ & $3.7^{+0.9}_{-1.2}$ & $4.1^{+0.9}_{-0.8}$ & $3.9\pm1.0$ \\
 $ N_\text{H,2}~(10^{22}\,\text{cm}^{-2})$ & --- & $17^{+13}_{-11}$ & --- & $27^{+20}_{-21}$ & $21^{+6}_{-7}$ & $19\pm8$ \\
 $ CF$ & --- & $0.36^{+0.42}_{-0.13}$ & --- & $0.21^{+0.31}_{-0.09}$ & $0.45^{+0.09}_{-0.08}$ & $0.42^{+0.16}_{-0.10}$ \\
 $ \mathcal{F}~(10^{-11}\,\text{erg}\,\text{cm}^{-2}\,\text{s}^{-1})\tablefootmark{a}$ & $8.99\pm0.18$ & $9.9\pm0.6$ & $7.7\pm0.5$ & $9.5^{+0.6}_{-0.5}$ & $10.49^{+0.44}_{-0.11}$ & $10.2^{+0.5}_{-0.4}$ \\
 $ A_\text{2}$ & --- & --- & --- & $\left(7.9^{+2.9}_{-2.5}\right)\times10^{-7}$ & --- & --- \\
 $ \Gamma$ & $1.65^{+0.06}_{-0.05}$ & $1.78\pm0.09$ & $1.48^{+0.11}_{-0.10}$ & $1.11\pm0.06$ & $1.82\pm0.06$ & $1.68^{+0.08}_{-0.07}$ \\
 $ E_\text{cut}~(\text{keV})$ & $7.1^{+0.6}_{-0.5}$ & $7.6\pm0.7$ & $5.5^{+1.0}_{-0.7}$ & --- & $<3.9$ & --- \\
 $ E_\text{fold}~(\text{keV})$ & $27.6^{+2.6}_{-2.2}$ & $33^{+5}_{-4}$ & $25.7^{+3.5}_{-2.8}$ & $9.0^{+0.8}_{-0.6}$ & $23.9^{+2.2}_{-1.8}$ & $27.3^{+3.0}_{-2.5}$ \\
 $ R_\text{bbody}~\text{(km)}$ & --- & --- & $0.23^{+0.07}_{-0.06}$ & --- & --- & --- \\
 $ kT~(\text{keV})$ & --- & --- & $1.88^{+0.21}_{-0.19}$ & --- & --- & --- \\
 $ \text{Eqw~FeK}\alpha~(\text{eV})$ & $18\pm12$ & $12^{+13}_{-12}$ & $18^{+12}_{-14}$ & $12^{+13}_{-12}$ & $6^{+10}_{-7}$ & $10^{+12}_{-10}$ \\
 $ CC_\text{B}$ & $1.003\pm0.008$ & $1.003\pm0.008$ & $1.003\pm0.008$ & $1.003\pm0.008$ & $1.003\pm0.008$ & $1.003\pm0.008$ \\
 $ CC_\text{XRT}$ & $0.62\pm0.06$ & $0.59\pm0.06$ & $0.59\pm0.06$ & $0.59\pm0.06$ & $0.58\pm0.06$ & $0.59\pm0.06$ \\
$\mathcal{L}~(10^{35}$\,erg\,s$^{-1}$)\tablefootmark{b} & $5.43\pm0.11$ & $6.0\pm0.4$ & $4.64\pm0.27$ & $5.76^{+0.32}_{-0.27}$ & $6.32^{+0.27}_{-0.07}$ & $6.14^{+0.27}_{-0.24}$ \\
\hline$\chi^2/\text{d.o.f.}$   & 717.03/660& 699.44/658& 694.78/658& 699.45/658& 763.48/658& 743.66/659\\$\chi^2_\text{red}$   & 1.086& 1.063& 1.056& 1.063& 1.160& 1.128\\\hline
\end{tabular}
\tablefoot{\tablefoottext{a}{unabsorbed powerlaw flux between 3--10\,keV}
\tablefoottext{b}{luminosity between 3--10\,keV for a distance of 7.1\,kpc}
}
\end{table*}

\begin{figure}
\begin{center}
\includegraphics[width=0.95\columnwidth]{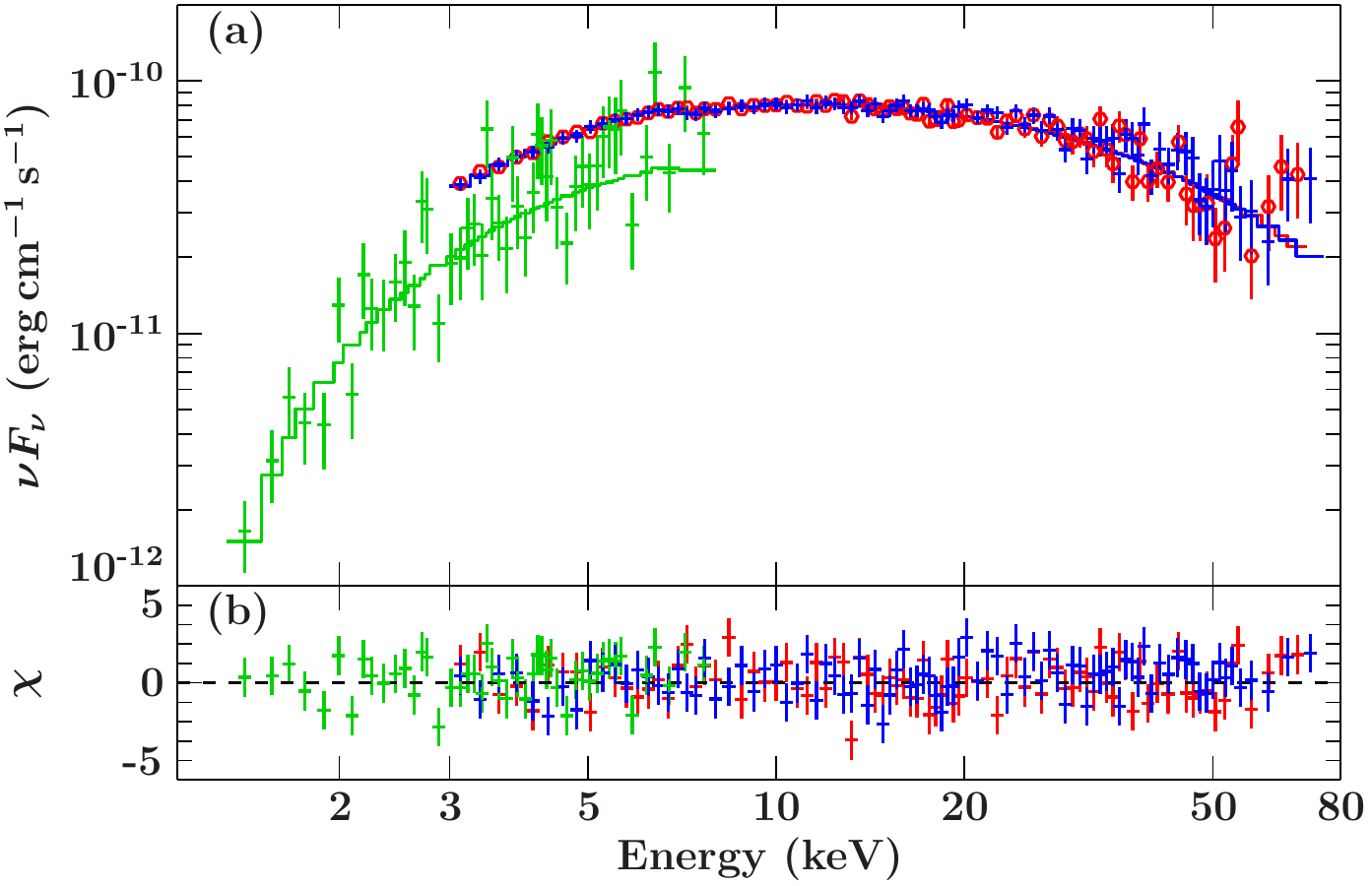}
\caption{{(a)} Unfolded \swift/XRT (green), \nustar/FPMA (red), and \nustar/FPMB (blue) phase-averaged spectra, together with the best-fit \texttt{highecut} model.
{(b)} Residual in terms of $\chi^2$ for the best-fit model. Data were strongly rebinned  for clarity.}
\label{fig:spec}
\end{center}
\end{figure}

While we cannot statistically distinguish between the \texttt{highecut} and \texttt{NPEX} model, we will base the remaining discussion on the \texttt{highecut} model as \citet{naik13a} found that it describes their data better. The \texttt{highecut} model was also successfully applied to  \exo  by \citet{wilson08a} and \cite{klochkov08b}, and  therefore allows a direct comparison with their results. Our results are all qualitatively the same for the \texttt{NPEX} model. Figure~\ref{fig:spec} shows the XRT and \nustar spectrum, together with the best-fit \texttt{highecut} model. 

We also tested a model consisting of a single absorbed \texttt{highecut} with an additional blackbody component at low energies for direct comparison with the results by \citet{reig99a}. This model results in the lowest \redchi\ value, but  statistically it is not significantly better than the partially covered version. 

Furthermore we searched for a neutral \feka line, modeled by a narrow Gaussian at 6.4\,keV, that has often been observed in this source \citep[e.g.,][]{reig99a,naik13a,ferrigno16a}. We do not find clear evidence of the presence of this line, with a 90\% upper limit on the equivalent width around 25--30\,eV, independent of the choice of the continuum model (Table~\ref{tab:phasavg}).

\subsection{Phase-resolved spectroscopy}

Following the discovery of a peculiar feature in the phase-resolved data of \xmm by \citet{ferrigno16a}, we extracted \nustar spectra in 100 phase-bins. We used a local period of 41.287054\,s (for more information about the timing solution, see Sect.~\ref{susec:pevolv}).
The \swift data did not provide enough signal to allow for phase-resolved analysis. 

The very high resolution in phase-space is possible thanks to \nustar's high throughput even at energies above 10\,keV. We assumed that neither the detector response nor the background changes on timescales  of the pulse period, and therefore used the respective phase-averaged  information.
We rebinned each spectrum to a \snr of 5 within ISIS and on average obtained signal up to 22\,keV  with about 60 bins per spectrum.

For spectral modeling, we used the partially absorbed \texttt{highecut} (without the blackbody component). We confirmed that using the \texttt{NPEX} model does not change our conclusions. Due to the reduced \snr, we had to simplify the spectral model and fixed the covering fraction, the absorption columns, the cutoff energy, and the folding energy at the respective values of the  phase-averaged fit. This approach left us with only two free parameters: the flux and the photon index ($\Gamma$). Instead of the photon index, we also tried  allowing one of the absorption parameters to vary, but  this resulted in large degeneracies and unconstrained parameters, due to the lack of coverage at very low energies. We therefore limited ourselves to capturing all observed spectral changes by a variable photon index. The absolute values of the photon index should consequently be taken with a grain of salt, but its variation is a very good tracer of spectral variability and changes in hardness.

\begin{figure}
\begin{center}
\includegraphics[width=0.95\columnwidth]{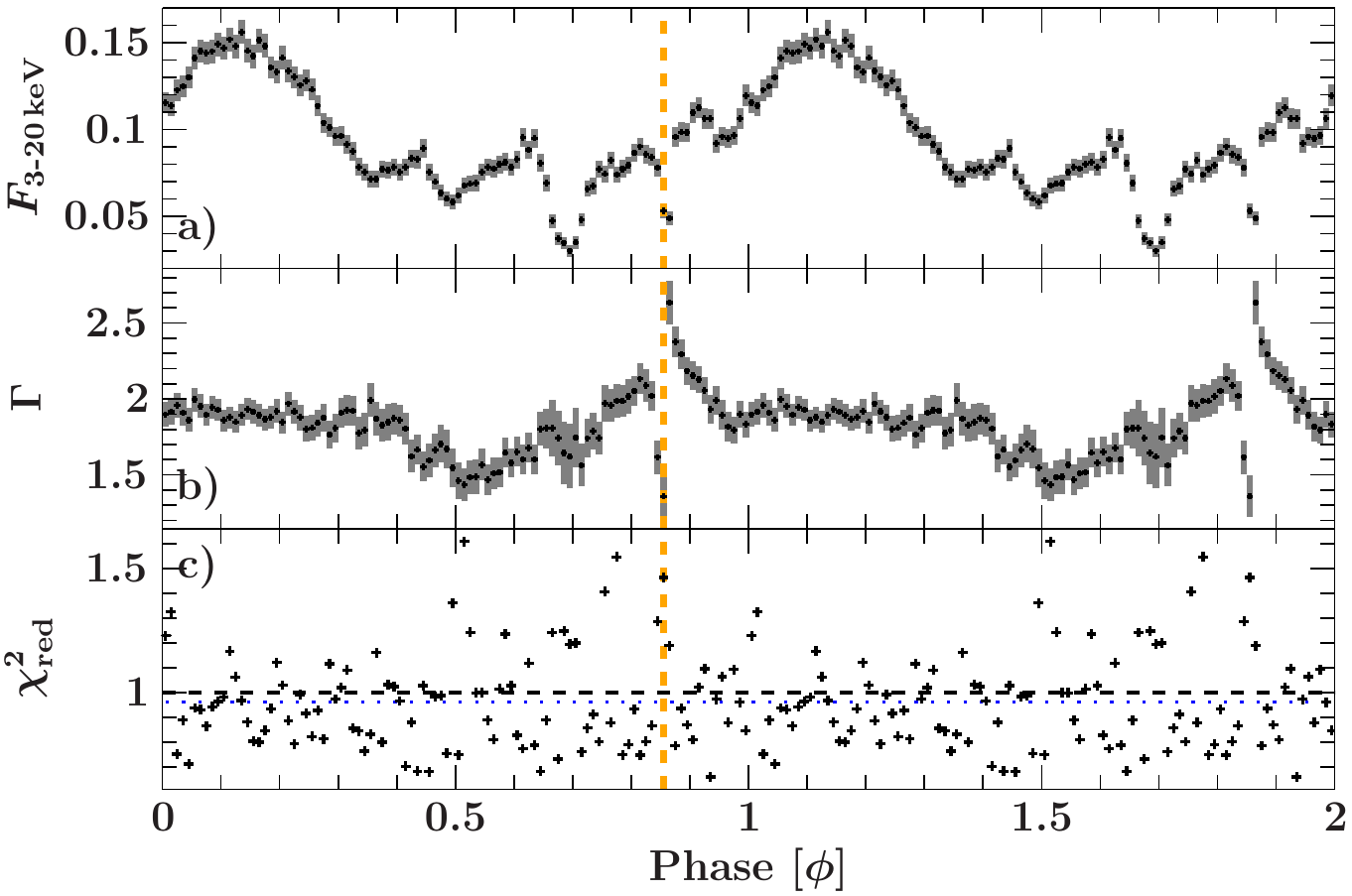}
\caption{Results of the phase-resolved spectral fits using 100 phase-bins with \nustar data only. {(a)} Flux between 3--20\,keV in keV\,cm$^{-2}$\,s$^{-1}$.
{(b)} Photon index. All other continuum parameters were frozen to the phase-averaged value. The orange dashed line indicates the  dip  to guide the eye.
{(c)} \redchi value for each phase-bin. The blue dotted line indicates the average \redchi overall bins.}
\label{fig:phasres}
\end{center}
\end{figure}

The results of this fit are shown in Fig.~\ref{fig:phasres}. Overall, the description of the data is statistically very good, with an average $\redchi$ of 0.96.
During most of the pulse, the photon index varies rather smoothly. In particular, no strong change is evident in the main peak of the pulse profile or during the low state around phase 0.7.

However, we can clearly make out an interesting feature around phase 0.85. The spectrum is significantly harder for about two phase-bins and then suddenly jumps from $\Gamma\approx1.5$ to $\Gamma>2.5$. This behavior is very similar to that observed by \citet{ferrigno16a}, i.e., a hardening followed by a sudden softening. The feature in the \nustar data is of a similar duration to that in the \xmm data, between 2--3\% of the pulse period. 

We also observe indications of a small phase shift between the feature in flux and in the photon index. The value of $\Gamma$ drops one phase-bin before the flux goes down. This phase shift seems to be much smaller in the \xmm data, if present at all \citep{ferrigno16a}.

\section{Long-term trend}
\label{sec:longterm}

\subsection{X-ray spectra}

\begin{figure*}
\begin{center}
\includegraphics[width=0.95\textwidth]{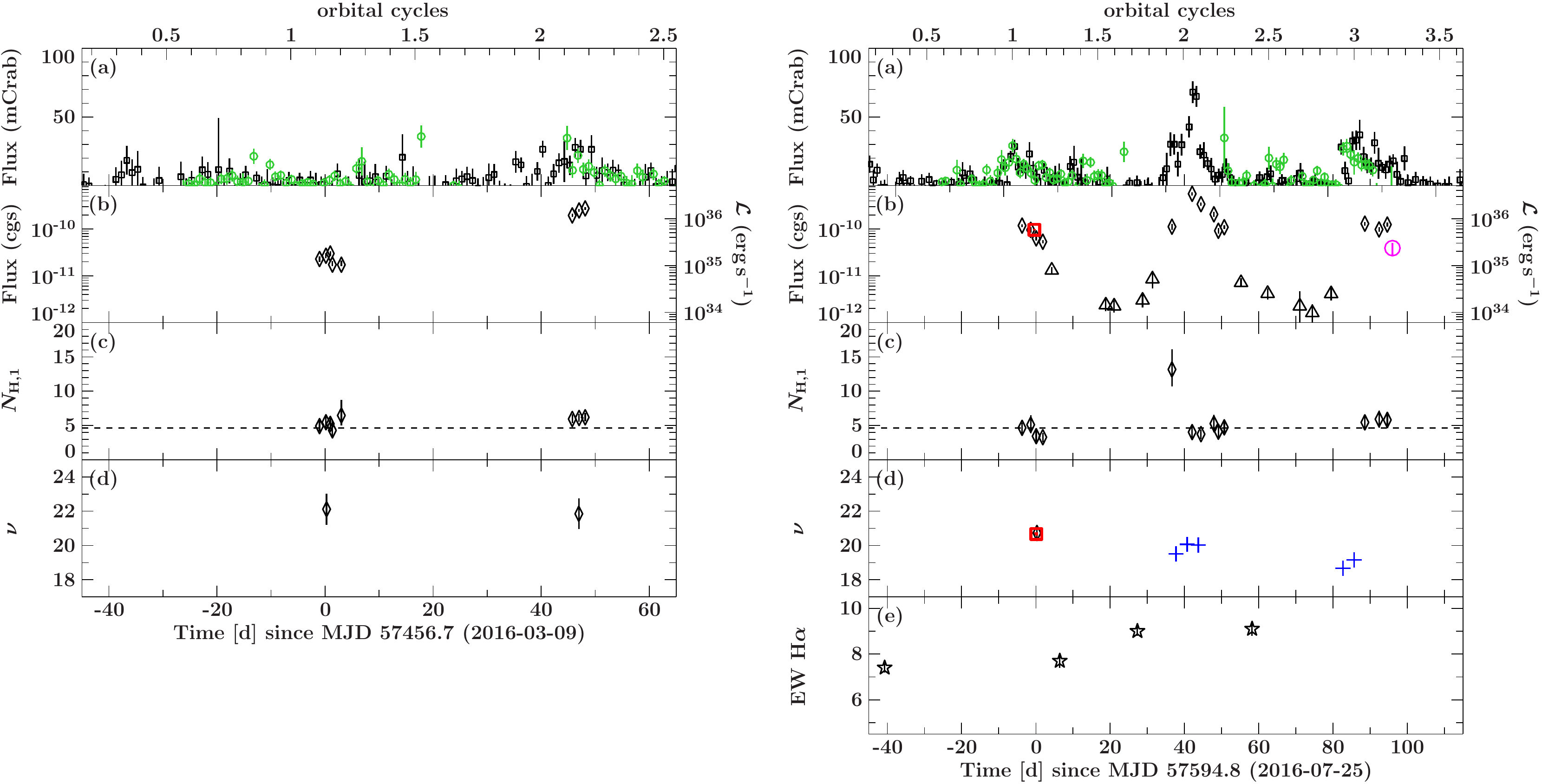}
\caption{{(a)} \swift/BAT (black) and MAXI (green) fluxes in mCrab from March to May 2016 (left) and July to October 2016 (right). The fluxes were rescaled to the respective mCrab count-rates, using 0.00022\,cts\,s$^{-1}$\,cm$^{-2}$ (15-50\,keV)  for BAT and 0.0033\,cts\,s$^{-1}$\,cm$^{-2}$ (2--20\,keV) for MAXI.
{(b)} Flux measured in the 3--10\,keV energy band in units of \ergcms. For the data points shown as triangles the absorption was frozen to $5.9\times10^{22}$\,cm$^{-2}$ as it could not be constrained due to the low S/N. Furthermore, no pulsation search could be performed on those data for the same reason.
The \nustar flux is shown by a red square and the \inte/ISGRI flux is shown by a magenta circle and measured between 25--40\,keV.
The $y$-axis is scaled logarithmically. The right hand $y$-scale gives the implied spherical luminosities for a distance of $d=7.1$\,kpc.
{(c)} Global absorption column $\nhone$ measured with \swift/XRT in  $10^{22}$\,cm$^{-2}$.  The columns of \nustar and ISGRI were tied to the closest XRT point and are therefore not shown.
{(d)} Pulse frequency $\nu - 24.2$\,mHz in $\mu$Hz. Blue crosses are values measured by \fermi/GBM, black diamonds are from \swift/XRT, and the red square is the  \nustar value.
{(e)} Equivalent width of the H$\alpha$ line, as measured in the NOT/ALFOSC spectra.
}
\label{fig:batlc}
\end{center}
\end{figure*}

We analyzed \swift/XRT snapshots taken before and after the \nustar observation and a serendipitous  \inte observation to put the \nustar observation into context. Our first \swift campaign was triggered following the repeated non-detections of \tone outbursts, and covered the periastron passages in March and April 2016. A second campaign between July and September 2016 covered almost two complete orbital cycles (Fig.~\ref{fig:batlc}). 

As the XRT data  do not provide enough information to constrain the continuum parameters at the same time as the absorption column, we fitted them all simultaneously with the \nustar and \inte/ISGRI data (which were taken during an outburst that occurred two orbital cycles after the \nustar data), requiring that all spectra have the same photon index, cutoff energy, and folding energy. 
In this approach, the \nustar data provide the main constraints on the continuum. The ISGRI data also probe the high-energy continuum,  albeit at a much lower \snr.  They are fully consistent with the \nustar data,  indicating that the continuum did not change significantly between the two outbursts. 

The XRT data, on the other hand, allow us to investigate the behavior of the absorption column, i.e., the parameter to which it is most sensitive. As both \nustar and ISGRI lack coverage at energies where the absorption is relevant, we tied their respective absorption columns to the one measured by the respective contemporaneous XRT data. 
Furthermore, for the very faint XRT spectra that  have less than 150 total counts, we  set the absorption parameters to the best-fit values shown in Table~\ref{tab:phasavg} as they could not be constrained from the data.

In our partial covering model, we have three parameters to describe the complex absorption: \nhone, \nhtwo, and the covering fraction. We allowed in turn each of these parameters to be independently fit for each \swift spectrum, while the other two were tied across all data sets (i.e., treated as global parameters as defined by \citealt{kuehnel16a}).  We find equally good fits when allowing either the global column, \nhone, or the covering fraction $f$  to vary. However, allowing only \nhtwo to vary results in a significantly worse fit.

In Fig.~\ref{fig:batlc}{c}, we show the result for a variable \nhone. As can be seen, the absorption column is relatively constant over different outbursts and fluxes. We only find one significant outlier at MJD~57631.4 where the column increases drastically to $\left(13.6^{+2.9}_{-2.5}\right)\times10^{22}$\,cm$^{-2}$.  The results for a variable covering fraction are qualitatively similar; there is a strong increase for the same observation to $f=0.91\pm0.06$, while on average we observe $\left<f\right>= 0.56\pm${0.13}.
This outlier observation took place around orbital phase 0.93, but no other XRT observation covers this exact orbital phase.

The XRT data show that even in March 2016, when neither BAT nor MAXI measured residual flux, the source had not turned off completely but was still showing a luminosity around   $\left(1.4\pm0.2\right)\times10^{35}$\,erg\,s$^{-1}$ between 3--10\,keV.  This luminosity was very likely dominated by accretion, as shown by the hard X-ray spectrum and the still remaining pulsations (see Sect.~\ref{susec:pevolv}).  

The lowest luminosity during the \swift campaign was observed on MJD 57669 with $L_\text{X}=\left(1.0^{+0.6}_{-0.5}\right)\times10^{34}$\,erg\,s$^{-1}$. However, the very low count-rate did not allow us to search for pulsations. The spectral shape is consistent with the other XRT spectra, but could not be individually constrained. 

To obtain a better understanding of the spectrum at low fluxes, we therefore combined all spectra with a 3--10\,keV flux less than $5\times10^{-11}$\,\ergcms\ (marked by triangles in Fig.~\ref{fig:batlc}) into an average low flux spectrum. We describe the spectrum with a model based on the best-fit \nustar model with a partially covered continuum with a high-energy cutoff.  As a first approach we froze all parameters except the flux to the best-fit \nustar values (see Table~\ref{tab:phasavg}), but we did  not find an accetable fit ($\redchi=1.24$ for 18 dof), which indicates that the spectrum changed significantly.

\begin{figure}
\begin{center}
\includegraphics[width=0.95\columnwidth]{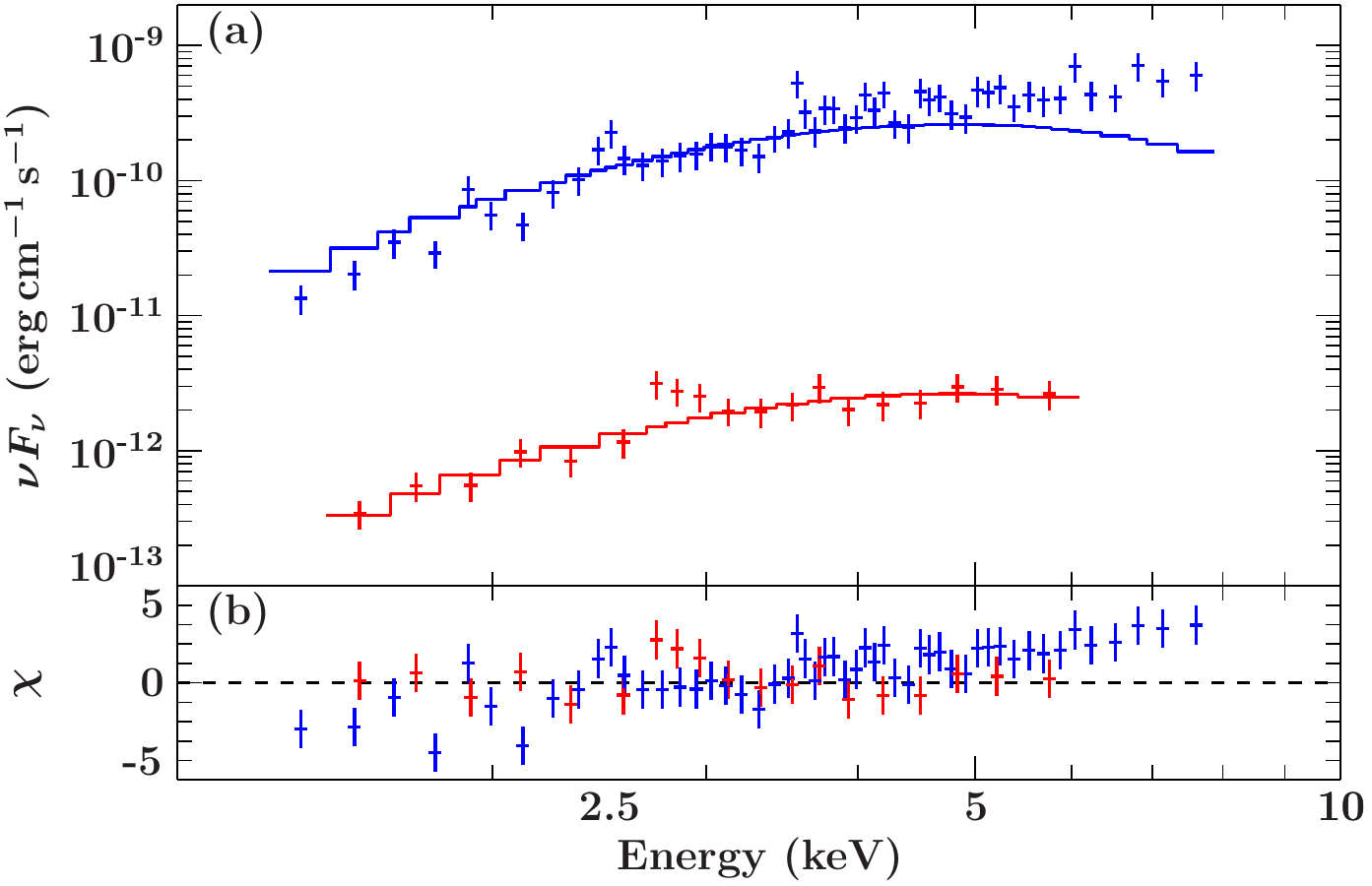}
\caption{ \swift/XRT spectra and residuals of the combined faint states (red) and during a single bright observation (ObsID 00030799031, blue). The model is the best-fit blackbody spectrum fitted to the faint state (see Table~\ref{tab:faint}), and only scaled in flux to match the brighter state. The brighter state is clearly harder than the faint data. }
\label{fig:faintspec}
\end{center}
\end{figure}

We then allowed either the photon index, one of the absorption columns, or the covering fraction to be variable. This resulted in an improved fit, with $\Delta \chi^2 \approx 6$ for each of those parameters (the second column of Table~\ref{tab:faint} shows the case with a variable photon index). We also tried models where pairs of these parameters were allowed to be variable.
With this approach, we found that we obtained a best fit when allowing the photon index and the global absorption column \nhone to vary. The best-fit parameters are presented in the third column of Table~\ref{tab:faint}.

\begin{table}
\caption{Best-fit model parameters for the faint XRT spectra.}
\label{tab:faint}
\begin{tabular}{rlll}
\hline\hline
Parameter & HighE 1 & HighE 2 &Bbody \\\hline
$ N_\text{H,1}~(10^{22}\,\text{cm}^{-2})$ & --- & $4.1^{+2.2}_{-1.8}$ & $1.1^{+1.4}_{-1.2}$ \\
$ \Gamma$ / $kT$~(keV) & --- &  $2.2^{+0.8}_{-0.7}$ & $1.22^{+0.31}_{-0.22}$ \\
$R$~(m) & --- & ---- &  $280^{+140}_{-90}$ \\
$\mathcal{L}~(10^{34}$\,erg\,s$^{-1}$)\tablefootmark{a} & $2.83\pm0.27$ & $2.4^{+0.5}_{-0.4}$ & $1.55^{+0.38}_{-0.30}$ \\\hline
$\chi^2/\text{d.o.f.}$   & 22.2 / 18 & 16.4 / 16 & 15.2 / 16 \\
\redchi & 1.24 &  1.03 & 0.95 \\\hline
\end{tabular}
\tablefoot{\tablefoottext{a}{luminosity between 3--10\,keV for a distance of 7.1\,kpc}}
\end{table}

We also tried to fit the data with just a thermal blackbody model, and find a slightly better fit (fourth column in Table~\ref{tab:faint}). We find a blackbody temperature of $kT=1.22^{+0.31}_{-0.22}$\,keV. This is somewhat hotter than the temperatures \citet{wijnands16a} and \citet{tsygankov16b} found in the quiescent phases of 4U\,0115+63 and V\,0332+53, but our luminosity is also slightly higher ($L_x = \left(2.2 \pm 0.4\right)\times10^{34}$\,erg\,s$^{-1}$). We find a radial size of the blackbody of $r=280^{+140}_{-90}$\,m, which is at the lower end compared to the values found by \citet{wijnands16a}.

\subsection{Pulse period }
\label{susec:pevolv}

The pulse period of \exo is normally well monitored by \fermi/GBM\footnote{\url{https://gammaray.msfc.nasa.gov/gbm/science/pulsars.html}} \citep{finger09a}. During the periastron passages between 2016 March and August  (MJD~57450 to MJD~57600), however, the source was too faint to be measurable by GBM and therefore the behavior of the spin-down was not captured. We therefore searched for periodicity in the \nustar and \swift/XRT data obtained during the periastron passages in that time frame. We transferred all time information to the solar barycenter, and corrected for the orbital Doppler shift using the ephemeris of \citet[Fit~2]{wilson08a}. We searched for a period around the known value of 41.3\,s in the \nustar source event lists, using the epochfolding technique \citep[e.g.,][]{leahy83a}. Uncertainties were estimated through a Monte Carlo simulation \citep{davies90a}. We find a period of $P_\text{NuSTAR} = 41.28705 \pm 0.00008$\,s ($1\sigma$ uncertainty) at epoch $t_0 =\text{MJD}~ 57594.37019$ and no evidence for a change in the period $\dot P$ over the observation.

To cross-check this value, we used the epoch folding technique on
the \swift-XRT light curves taken between MJD~57591
and MJD~57599.
However, in order to avoid
strong secondary maxima within the epoch folding result, we split
the light curves into segments where gaps in the data resulting
from bad time intervals are present. The resulting pulse periods
for all segments and observations are consistent with $41.22\pm0.13$\,s.

In order to further increase the precision of the XRT pulse period
measurement, we phase connected the pulse profiles of all light
curve segments using 16 phase
bins \citep{manchester77a,deeter81a}. The resulting phase shift,
$\delta \phi$, between each adjacent light curve segment is
degenerate by an integer multiple corresponding to the error propagated
uncertainty of the pulse period determined by epoch folding. This uncertainty
results in a number of possible pulse phase evolutions over the time
of the \swift observations. 

We consequently fitted each evolution assuming a
constant spin-change, i.e., $\delta \phi(t-t_0) = \nu t + 0.5\dot{\nu}
t^2 $ with the time of each observation $t$, a reference time
$t_0 = \mathrm{MJD}~57594.85875$, the pulse frequency $\nu = 1/P$,
and its derivative $\dot{\nu} = -\dot{P} / P^2$, where $\dot P$ is the pulse
period derivative. After having fitted all possible pulse phase evolutions
resulting from the mentioned degeneracy, we selected the evolution
providing the best goodness of fit ($\Delta \chi^2$ = 3.5 with 6 degrees
of freedom), which results in a final pulse period of
$41.2869\pm0.0004$\,s, in very good agreement with \nustar. 
Our best-fit period derivative, $\dot{P}$, is consistent with
zero, which is due to the short time range covered here.

We used a similar approach for the previous two periastron passages where we also had \swift data available (around MJD~57451 and 57493). As the source was overall much fainter, the uncertainties on the obtained periods were higher. We find  $41.2846\pm0.0008$\,s on MJD~57457.0 and $41.2850\pm0.0008$\,s on MJD~57503.7, again with no evidence for $\dot P$ within each epoch.

Folding the  \swift/XRT data taken during apastron did not result in a significant detection of pulsations at the 99\% level when taking the number of trials into account \citep[based on the $L$-statistic, ][]{davies90a}. Given the low photon counts in these data, the non-detections are not surprising and do not necessarily imply that the source is not pulsing. From a rough estimate of the detectable pulsed fraction \citep{brazier94a}, we find that the data are only sensitive to pulsed fractions $>$75\%, while the pulsed fraction during the periastron observations is only $\sim$40\% in XRT.

\begin{figure}
\begin{center}
\includegraphics[width=0.95\columnwidth]{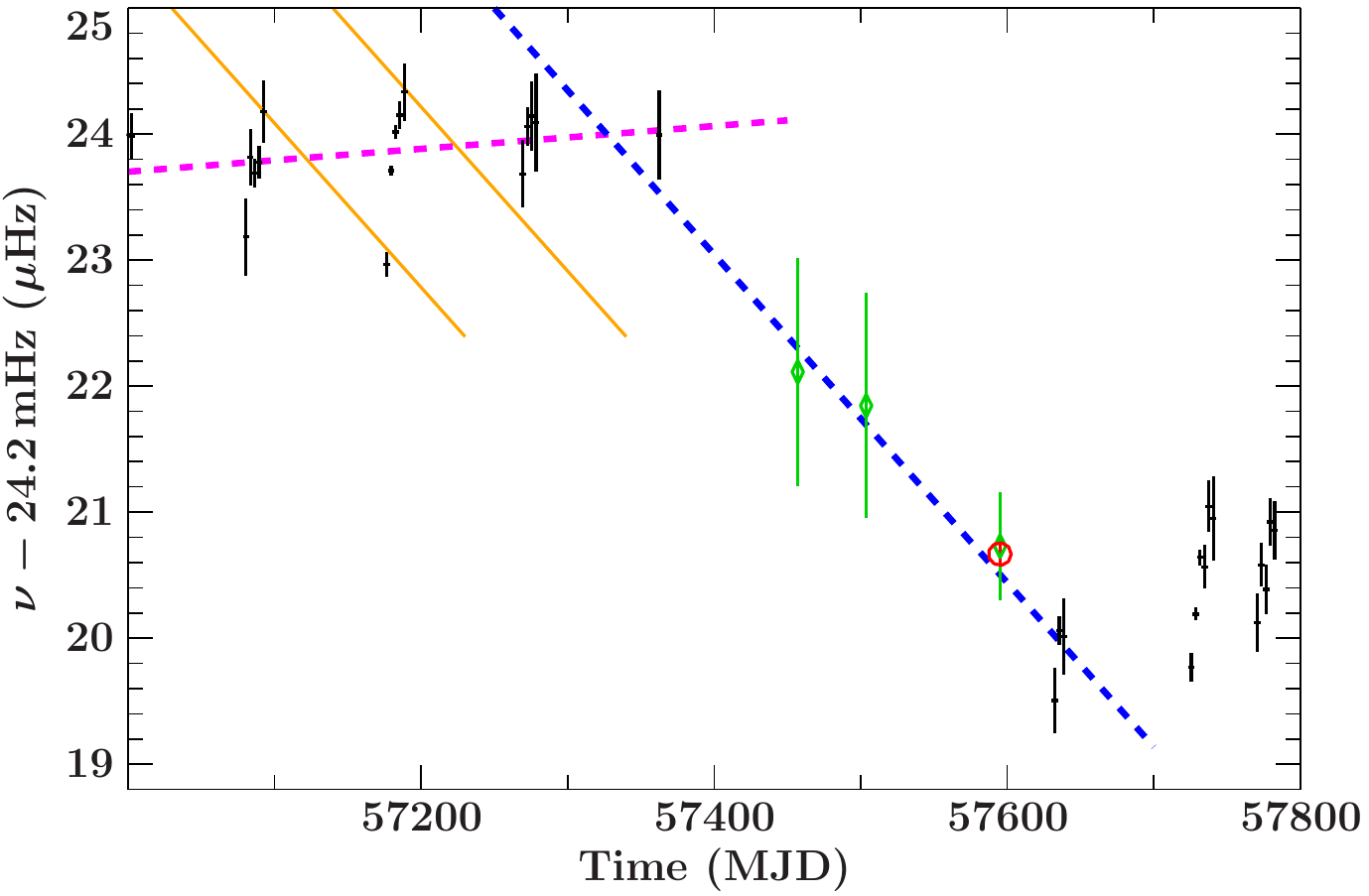}
\caption{Pulse frequency evolution of \exo, as measured with \fermi/GBM. The \nustar measurement is shown as a red circle, the \swift measurements are shown as green diamonds. Each group of black data points corresponds to one \tone outburst visible to GBM. During each outburst, the source spins up, while between outbursts it spins down again. With the start of the quiescence period, a continuous spin-down started.
The  dashed lines are linear regression fits to  the active phase, which shows a secular spin-up trend (magenta), and to the spin-down in quiescence (blue), respectively. The orange solid lines have the same slope as the best-fit spin-down linear regression to guide the eye between the spin-up episodes during outbursts. The data after MJD~57700.0 show that with the again stronger outbursts, \exo is also starting to spin-up again. }
\label{fig:perevol}
\end{center}
\end{figure}

The \nustar and \swift results are put into context with the \fermi/GBM results in Fig.~\ref{fig:perevol}. As can be seen, our results confirm an almost linear long-term spin-down with a rate of change of $\dot{P}=\left(2.7814\pm0.0003\right)\times10^{-10}$\,s\,s$^{-1}$. 
This spin-down is of a comparable strength to the one typically observed between outbursts (see Fig.~\ref{fig:perevol}). This indicates that the accretion rate was very low during the whole quiescent epoch and likely not higher than during typical apastron passages.

\subsection{Optical monitoring}

\begin{figure}
\begin{center}
\includegraphics[width=0.95\columnwidth]{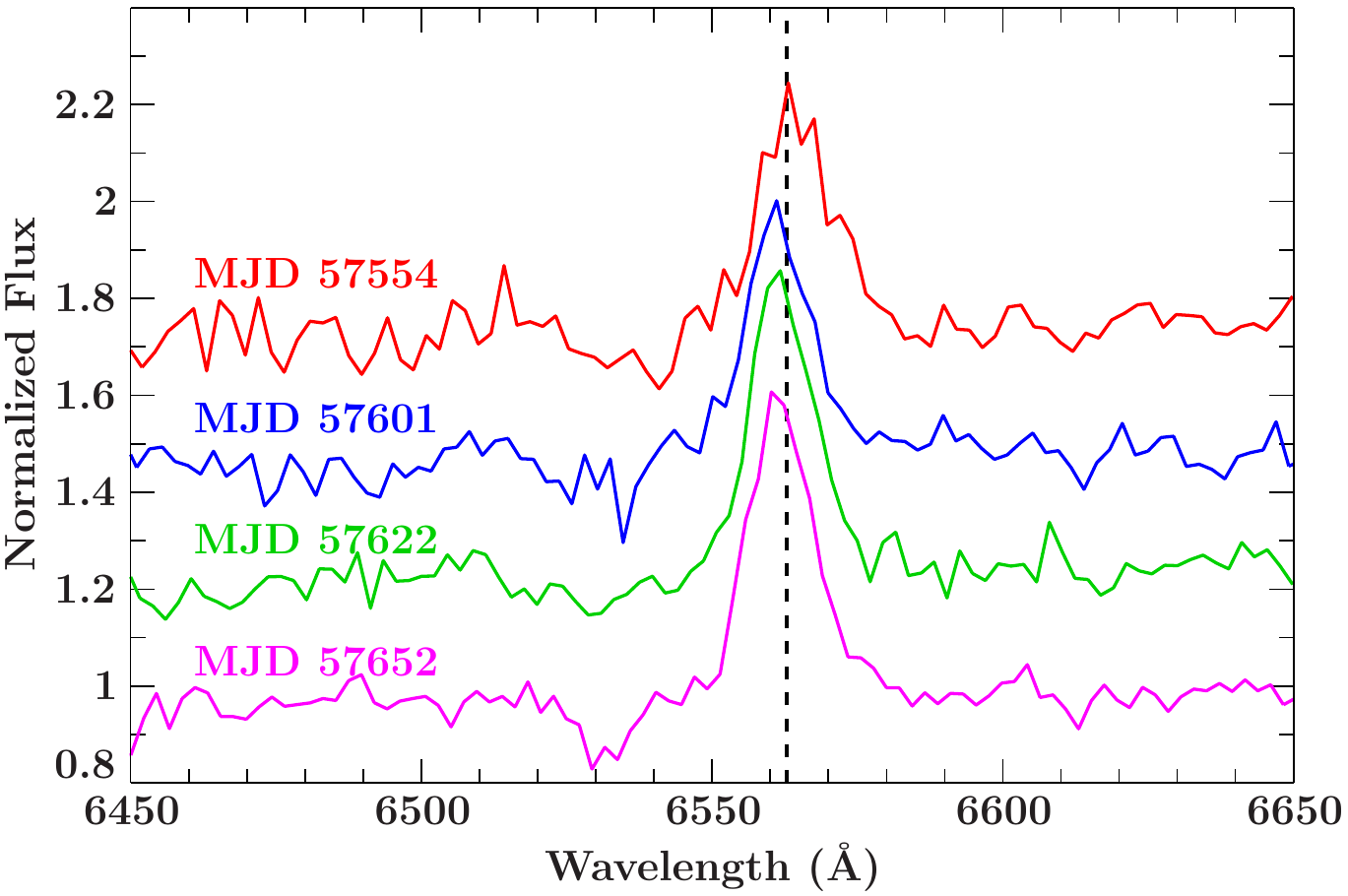}
\caption{\ha line of \exo as measured with the NOT/ALFOSC on four different nights between June and September 2016. The spectra are shifted in steps of 0.25 in flux for visual clarity. The dashed line indicates the rest wavelength of \ha.}
\label{fig:halpha}
\end{center}
\end{figure}

Between June and September  2016, we obtained four measurements of the \ha line with the NOT to trace the evolution of the circumstellar disk. 
We show the shape of the \ha line in Fig.~\ref{fig:halpha}. 
We measure equivalent widths (EWs) between 7--9\,\AA\xspace in all four spectra and show their evolution in Fig.~\ref{fig:batlc}e. 
Similar values were measured by \citet{reig98a} and \citet{wilson02a}, just after the period when there were also missing X-ray outbursts (MJD $\lesssim$50000), and coinciding with a change in orbital phase of the peak of the outbursts \citep{laplace17a}. 
These events, and the ones we present here, are separated by about 21\,yr, roughly the same separation as for the \ttwo outbursts. 

These similar periodicities could be explained by  Kozai-Lidov oscillations with a 21-year period \citep{laplace17a}. The Kozai-Lidov  effect \citep{kozai62a, lidov62a} causes an initially misaligned accretion disk to oscillate between a larger eccentricity and a larger inclination with respect to the orbital plane. As the observed EW is directly connected to the inclination of the circumstellar disk with the line of sight,  the periodic occurrence of low EWs, coincident with a low X-ray state,  could be explained by a change in inclination on the Kozai-Lidov timescale. The EW therefore does not necessarily trace the size or density of the disk, but only its projected area. As \citet{laplace17a} argue, it is likely that the faint-states and the change in outburst phase are connected with a highly inclined disk (with respect to the orbital plane), which is observed as a low EW of the \ha line. On the other hand, the giant outbursts coincide with a high eccentricity of the disk and a low inclination, resulting in a much larger EW.

We note, however, that \citet{baykal08a} also measured  low values of the EWs after the giant \ttwo outburst in 2006 June. As the Kozai-Lidov oscillation would predict a lower inclination and therefore higher EW of the \ha line during that time, this measurement might be instead explained by a disk-loss produced by the large quantity of accreted material.

The line profiles we observe show marginal indications for a P Cygni profile, with an estimated velocity of $\sim$1300\,km\,s$^{-1}$. If this profile were real, it would indicate the presence of a strong stellar wind, which would be responsible for the observed absorption. However, the low \snr of the data prevents us from drawing any further conclusions.

\section{Discussion}
\label{sec:disc}

\subsection{Spectral shape}

We find that a partially covered \texttt{highecut} model describes the phase-averaged \nustar and contemporaneous XRT spectrum very well (Table~\ref{tab:phasavg}). 
A similar model has traditionally been  used to describe the spectrum of \exo, e.g., by \citet{reynolds93a} and \citet{wilson08a}. 
All these works have found a consistent correlation between luminosity and photon index, i.e., softer spectra at higher luminosities. \citet{reig99a} find that this correlation levels off at lower luminosities around $10^{36}$\,erg\,s$^{-1}$. In our data, however, we find a much softer photon index than \citet{reig99a} ($\Gamma = 1.81^{+0.08}_{-0.10}$), thereby reversing the typical correlation. This is somewhat reminiscent of the behavior found in black hole binaries, which in the hard state show the hardest photo index, typically around 1\% of the Eddington luminosity, and soften significantly towards lower and higher fluxes \citep[e.g.,][]{tomsick01a,yang15a}.
Depending on the bolometric corrections and mass of the neutron star, the lowest luminosities observed by \citet{reig99a} of around $2\times10^{36}$\,erg\,s$^{-1}$ correspond to about 1\% Eddington.

On the other hand, \citet{wilson08a} observed a relatively stable cutoff  energy around 12.5\,keV until the end of the outburst, where it dropped to values around 10\,keV at the lowest fluxes. In our data we find an even lower cutoff energy of $6.8 \pm 0.5$\,keV (Table~\ref{tab:phasavg}), which seems to agree roughly with the trend found by \citet{wilson08a} towards the end of the outburst and therefore does not require a different behavior at very low luminosities. 

\citet{klochkov07a} also use  a  \texttt{highecut} to describe \inte data taken during the 2006 giant outburst. They find an even higher cutoff energy around 20--25\,keV, depending on the exact model. 
The main difference between the \inte spectral shape and the one we find at low luminosities with \nustar is the curvature between $\sim$5--30\,keV, which is much less pronounced in our data. This could be related to the possible presence of a cyclotron resonant scattering feature (CRSF) or  a 10\,keV bump in the brighter \inte data \citep{klochkov07a}.
\citet{wilson08a}, on the other hand, find evidence for an absorption feature around 10\,keV, which they interpret as a CRSF. 

We tried to include a CRSF or 10\,keV bump, but do not find any evidence for such a feature in that energy range. If we assume a CRSF with the energy and width fixed to the values found by \citet[$E=11.44$\,keV and $\sigma=3.08$\,keV]{wilson08a} we can put a stringent upper limit on the optical depth\footnote{$\tau$ is total optical depth of the line and not the line strength $d$, as defined in the \texttt{gabs} model. They are connected via $\tau = d/(\sigma\sqrt{2\pi})$.}  $\tau \leq 0.012$, which is about one order of magnitude below the best-fit value found by \citet{wilson08a}.

\citet{klochkov08b} measured a CRSF with \inte at 64\,keV during a giant outburst. While the source is detected in the phase-averaged \nustar data up to $\sim$70\,keV, we do not find evidence for this CRSF.  If we assume an energy of 64\,keV and a width of 5\,keV, we can put an upper limit on the optical depth $\tau <0.07$. This is significantly smaller than  the typically observed depths (e.g., the harmonic line in Vela~X-1 at 55\,keV with $\tau\approx0.3$, \citealt{velax1nustar}). However,  as we do not have coverage above the proposed CRSF energy, it is still possible that the line is subsumed in our continuum model. As the \swift monitoring data suggest a high magnetic field, it is also possible that the CRSF energy moved out of the observable \nustar range with the lower luminosity \citep[for a possible luminosity dependence of the CRSF energy see, e.g.,][]{becker12a}.

\subsection{Propeller effect and thermal emission}
Previous studies \citep{reynolds96a, wilson02a} have postulated that the propeller effect might occur at luminosities below $10^{36}$\,erg\,s$^{-1}$. This effect would shut off most accretion and result in a strong drop in the luminosity.
However, in the \nustar and XRT data we observe pulsations and a hard spectrum at luminosities as low as $1.4\times10^{35}$\,erg\,s$^{-1}$ indicating that accretion is still ongoing and channeled by the $B$-field. Our results therefore reduce the upper limit for the propeller effect by almost an order of magnitude.

Analyzing low-luminosity data of A\,0535+26, \citet{rothschild13a} find that in this source pulsations are sometimes present at luminosities as low as $2\times10^{34}$\,erg\,s$^{-1}$. As the postulated cyclotron line energies and therefore magnetic fields in the two sources are similar, we might expect to observe pulsations in \exo at even lower luminosities.
These low luminosities are sustainable by just the wind-loss rate of the Be companion, even if no accretion from the circumstellar disk occurs.

The  lowest luminosities we observed were $\left(1.0^{+0.6}_{-0.5}\right)\times10^{34}$\,erg\,s$^{-1}$. However, at these low fluxes, the data are not statistically sensitive to pulsations at the typical observed level, and we therefore cannot determine if they are still present. However, by averaging over all low flux observations we find that the spectrum is very soft and well described by a purely blackbody continuum with a temperature of  $kT=1.22^{+0.31}_{-0.22}$\,keV. This softening could indicate that the source has actually entered the propeller regime and we are only observing the neutron star surface (or only the hot spots at the magnetic poles) that are cooling.  This idea has also been discussed by \citet{wijnands16a} and \citet{tsygankov16b} for low-luminosity observations of V\,0332+53 and 4U\,0115+63 and by  \citet{reig14a} for SAX~J2103.5+454.

We note that the unabsorbed continuum flux before each periastron passage is slowly rising, while still being relatively low. A rising flux, however, cannot be explained by neutron star cooling, and we therefore infer that some accretion power needs to contribute to the observed luminosity. Furthermore, the decline curve does not seem to follow the expected cooling curve of the neutron star surface because it is also likely influenced by residual accretion. Nonetheless, we find strong indications that the spectral shape is significantly softer and that the luminosity seems to  drop to the low levels rather abruptly. 

If we assume that accretion has stopped during the low flux state, the onset of the propeller effect happens somewhere between $2\times10^{34}$\,erg\,s$^{-1}$ and $1\times10^{35}$\,erg\,s$^{-1}$. 
The propeller effect takes place when the magnetospheric radius is larger than the corotation radius \citep{illarionov75a, cui97a, campana02a}, resulting in a critical luminosity $L_\text{prop}$ of
\begin{equation}
L_\text{prop} = 7.3 \times k^{7/2}  P^{-7/3}  R_6^5  B_{12}^2  M_{1.4}^{-2/3} \times 10^{37}\,\text{erg}\,\text{s}^{-1}
,\end{equation}
where $B_{12}$ is the magnetic field strength in $10^{12}$\,G, $P$ is the pulse period in seconds, $M_{1.4}$ is the neutron star mass in 1.4\,\msun, and $R_6$ is the neutron star radius in $10^6$\,cm. The factor $k$ parameterizes the effective coupling radius of the accretion disk to the magnetic field, which we set to 1 here \citep{wang96a}. For a derivation of this expression see the Appendix.

The critical luminosity obviously depends strongly on the magnetic field strength.
 Assuming a magnetic field strength of around $5\times10^{12}$\,G at the pole, as inferred from the 64\,keV CRSF measured by \citet{klochkov07a},  {translates to a field of about $2.5\times10^{12}$\,G at the equator, where the accretion disk interacts with the magnetic field}. This field strength would put the onset of the propeller effect around {$8.8\times10^{34}$\,erg\,s$^{-1}$, exactly in the expected range}. 
  However, if we assume $k=0.5$, which is often done for simple accretion disks \citep[e.g.,][]{ghosh79a}, {we find $L_\text{prop}  \approx 7 \times10^{33}$\,erg\,s$^{-1}$, which is below the lowest flux data of \swift.}

If we assume the much lower field of $1\times10^{12}$\,G postulated by \citet{wilson08a}, the luminosity for the onset of the propeller effect would drop to below {$2\times10^{33}$\,erg\,s$^{-1}$ in the spherical case and to $\approx 2\times10^{32}$ for $k=0.5$}. 
Such a low field seems therefore unlikely to be consistent with our observations.

However, we note that the calculation of the propeller luminosity is strongly dependent on the accretion flow, and therefore contains a rather large systematic uncertainty (see also the Appendix). In particular the calculation is based on spherical accretion, which is not necessarily a good approximation for \exo. Any value should therefore be taken with caution, but the rough estimate results in values consistent with the inferred luminosity for the onset of the propeller effect from the \swift/XRT data.

Only in the higher magnetic field case can \exo enter the propeller regime at all, according to the model presented by \citet{tsygankov17a}. In this model a cool, nearly unionized accretion disk can form for sufficiently slow rotators and weak magnetic field strengths. This disk will penetrate deeper into  the magnetosphere and can sustain persistent accretion at a low level. Following Eq. 7 of \citet{tsygankov17a} we find that the neutron star would have to spin with a period of 20\,s or faster to enter the propeller regime assuming a field of $\sim$$1\times10^{12}$\,G, but this is well below the critical period for all realistic values of $k$ in the high magnetic field case. This again shows that our observations are more consistent with a high magnetic field.

\subsection{Accretion geometry}

The phase-resolved analysis revealed a feature similar to that observed in \xmm, during which the photon index rapidly changes for about 3\% of the pulse period.
\citet{ferrigno16a} interpreted this feature as part of the accretion column passing through the line of sight, resulting in a significant rise in absorption column and  in a harder spectrum, due to more reprocessing of the emerging radiation. With \nustar, we are not able to probe variations in \nh, due to the limited coverage at low energies. \citet{ferrigno16a} further argue that the hotspot size should scale with luminosity as $\phi \propto L^{1/7}$, which would result in a smaller hotspot in the \nustar observations. However, we do not observe this effect; instead, the duration of the feature in \nustar seems to be {similar to or longer  than that in \xmm}. This indicates either that the accretion column size is not following the expectations or that the feature is independent of the accretion column. 

A duration of 2\% implies a size of around 1.3\,km at the equator of the neutron star (assuming a canonical radius of 10\,km). However, as the accretion column is likely at higher latitudes than the equator, this number can only be regarded as an upper limit. For a detailed interpretation, other effects (e.g., light-bending due to the strong gravity in the vicinity of the neutron star) have to be taken into account, which are beyond the scope of this paper.

During the brighter observations of the \swift/XRT monitoring we find a very stable spectrum, with an average absorption column of \nhone = $6.1\times10^{22}$\,cm$^{-2}$. Only one XRT observation shows a significantly enhanced \nh value. This is similar to the behavior seen in GX~304$-$1, which also showed enhanced absorption close to the periastron passage in some orbits 
\citep{rothschild17a}. This was interpreted as a warped and precessing Be-disk passing through the line of sight \citep{kuehnel17a}. However, the available \exo data do not allow us to put constraints on the precession frequency or size of the Be-disk;  these parameters are likely different from the long-term oscillation due to the Kozai-Lidov effect discussed by \citet{laplace17a}, as those timescales are much longer.  From optical monitoring we can rule out a drastically increased Be-disk, as the equivalent width of the H$\alpha$ line is constantly low.

\section{Conclusion}
\label{sec:concl}
We have presented an analysis of \nustar, \swift, and \inte data, supported by optical spectra with the NOT, and taken at very low luminosities of \exo. These data provide an in-depth look into the low-luminosity regime of this accreting neutron star. The \nustar data were taken at a luminosity around $\left(1.4\pm0.2\right)\times10^{35}$\,erg\,s$^{-1}$ in March 2016 (MJD~57457). In these data we still observe a hard spectrum and pulsations, clear evidence of ongoing accretion. We confirm the sudden change in spectral hardness found over a narrow range of pulse phases in the \xmm data in phase-resolved spectroscopy \citep{ferrigno16a}, indicating that the physical reason for this feature is independent of luminosity. More detailed investigations at more luminosity levels are necessary to understand its origin;  correct modeling of the relativistic effects around the neutron star is also necessary \cite[see, e.g.,][]{kraus03a, falkner16a}.

The lowest overall observed luminosity was around $2\times10^{34}$\,erg\,s$^{-1}$ (August 2016) with \swift/XRT. In this observation the spectrum was much softer. We interpret this as the onset of the propeller effect at luminosities just below $10^{35}$\,erg\,s$^{-1}$, which is in agreement with a strong magnetic field of $5\times10^{12}$\,G or more, as proposed by \citet{klochkov08b}.
We cannot constrain the spectral shape well at the lowest luminosities,  but find that it is well described by a thermal spectrum with a temperature of around 1.2\,keV. We interpret this spectrum as possibly originating from the cooling surface of the neutron star. As \exo has now picked up its regular outburst behavior again, further observations at these low luminosities are difficult. Nonetheless, with the expected 21-year periodic behavior \citep{laplace17a}, another low-activity state might occur during the lifetime of \textsl{Athena}, which will also provide the necessary sensitivity to study the apastron spectrum in detail.

\hspace{\baselineskip}

\begin{acknowledgements}
We thank the referee for the useful comments that helped to improve the paper.
We would like to thank the \nustar PI, Fiona Harrison, for accepting our observations in Director's Discretionary Time, and  the schedulers and the SOCs of \swift and \nustar for making them possible.
We are grateful for the support Neil Gehrels showed us during this and many other projects. May he rest in peace.
FF  is supported by a European Space Agency (ESA) Research Fellowship at the European Space Astronomy Centre (ESAC), in Madrid, Spain. 
JJEK acknowledges support from the Academy of Finland grants 268740 and 295114 and the ESA research fellowship program.
This work was supported under NASA Contract No. NNG08FD60C, and
made use of data from the {\it NuSTAR} mission, a project led by
the California Institute of Technology, managed by the Jet Propulsion
Laboratory, and funded by the National Aeronautics and Space
Administration. 
This research has made use of the {\it NuSTAR}
Data Analysis Software (NuSTARDAS) jointly developed by the ASI
Science Data Center (ASDC, Italy) and the California Institute of
Technology (USA). 
This work made use of data supplied by the UK Swift Science Data Centre at the University of Leicester.
We would like to thank John E. Davis for the \texttt{slxfig} module, which was used to produce all the figures in this work.
This research has made use of MAXI data provided by RIKEN, JAXA, and the MAXI team.
The \swift/BAT transient monitor results were
provided by the \swift/BAT team.
The data presented here were obtained in part with ALFOSC, which is provided by the Instituto de Astrof\'isica de Andalucia (IAA) under a joint agreement with the University of Copenhagen and NOTSA.
IRAF is distributed by the National Optical Astronomy Observatories, which are operated by the Association of Universities for Research in Astronomy, Inc., under cooperative agreement with the National Science Foundation.
\end{acknowledgements}

\begin{appendix} 
\section{Calculation of the propeller luminosity}
We based our calculations on the seminal works by \citet{lamb73a}, \citet{elsner77a}, and \citet{ghosh79a}, and the corresponding assumptions therein. In particular,  we assume spherical accretion with the velocity $v(r)$ at the magnetospheric radius being equal to the freefall velocity $v_\text{ff}(r)$:
\begin{equation}
v(r) = v_\text{ff}(r) =  \sqrt { 2GM / r }
.\end{equation}
Here, $M$ is the mass of the neutron star and $G$ is the gravitational constant. 
The magnetospheric radius is defined as the radius  $r_m$ at which the pressure of the magnetic field $B$ becomes similar to the pressure of the infalling matter with density $\rho$:
\begin{equation}
\frac{B(r_m)^2} { 8 \pi} = \rho(r_m)v(r_m)^2
.\end{equation}
It should be noted that we assume steady accretion, as is commonly done in the literature, and not accretion driven through Rayleigh-Taylor instabilities \citep[see discussion in][]{elsner77a}. The latter case, with magnetic pressure expressed as $B(r_m)^2/ { 4 \pi}$, would result in a factor of two larger luminosities, which is well within the systematic uncertainties of this estimate.

Using the continuity equation\footnote{assuming a cross section of $4\pi$ and neglecting a possible reduction by a factor $\xi$ for non-spherical accretion \citep{lamb73a}} $\dot M = 4\pi  \rho  r^2 v(r)$ and the expression of the magnetic field strength at the magnetospheric radius $B = \mu/ r_m^3$, we can write
\begin{equation}
\frac{\mu^2}{8\pi r_m^6} = \frac{\dot M}{4\pi r_m^2 v(r_m)} v(r_m)^2  =  \frac{\dot M}{4\pi r_m^2} \sqrt{\frac{2GM}{r_m}}
.\end{equation}
Solving for $r_m$ we find
\begin{equation}
r_m = \dot M^{-2/7} \mu^{4/7} (GM)^{-1/7} 2^{-3/7}
.\end{equation}

Replacing the magnetic moment $\mu$ with the surface magnetic field strength $B_\text{NS} = \mu / R_\text{NS}^3$ at the magnetospheric equator with the neutron star radius $R_\text{NS}$ and substituting the luminosity $L = \dot M GM / R_\text{NS}$, we obtain
\begin{equation}
r_m = B_\text{NS}^{4/7} R_\text{NS}^{12/7} \left(GM\right)^{-1/7}  2^{-3/7}\left({\frac{L R_\text{NS}}{GM}}\right)^{-2/7}. \\
\end{equation}

The correct inner radius $r_0$ of the disk depends on how the magnetic field threads the disk, which in turn depends strongly on the assumed accretion flow \citep[see, e.g.,][and references therein]{wang96a}. Here we approximate this factor with a scalar factor $k$, with $k$ typically between 0.5--1:
\begin{equation}
r_0 = k \times r_m
.\end{equation}

Replacing the variables with values of the expected order of magnitude, i.e., $L_{37}$ is the luminosity in $10^{37}$\,erg\,s$^{-1}$\,cm$^{-2}$, $B_{12}$ is the magnetic field in $10^{12}$\,G, $M_{1.4}$ is the mass of the neutron star in $1.4\,M_\odot$, and $R_6$ is  the neutron star in  $10^6$\,cm, we write
 \begin{equation}
r_0 = 3.0\times10^{8} k  B_{12}^{4/7} M_{1.4}^{1/7}L_{37}^{-2/7} R_6^{10/7}  \,\text{cm}
.\end{equation}

On the other hand, the Keplerian co-rotation radius is defined by
\begin{equation}
r_c = (GM)^{1/3} \left(\frac{P}{2\pi}\right)^{2/3} = 1.7\times 10^8 P^{2/3} M_{1.4}^{1/3} \,\text{cm}
\end{equation}
with $P$ being the pulse period.

The onset of the propeller effect occurs at a luminosity $L_\text{prop}$, where $r_m = r_c$ and thus

\begin{equation}
1.7\times 10^8 P^{2/3} M_{1.4}^{1/3} \,\text{cm} = 3.0\times10^{8} k B_{12}^{4/7} M_{1.4}^{1/7}L_\text{prop}^{-2/7} R_6^{10/7}  \,\text{cm}
,\end{equation}which finally gives us 
\begin{equation}
L_\text{prop} = 7.3 \times k^{7/2} P^{-7/3} R_6^{5} B_{12}^2 M_{1.4}^{-2/3}
\end{equation}
in units of $10^{37}$\,erg\,s$^{-1}$.
\end{appendix}

\end{document}